\documentclass[amsmath,amssymb,showpacs,preprint]{revtex4}
\usepackage{epsfig}
\usepackage{graphicx}

\begin{document}

\title{$\Lambda$CDM model with a scalar perturbation vs. preferred direction of the universe}

\author{Xin Li}
\email{lixin@ihep.ac.cn}
\author{Hai-Nan Lin}
\email{linhn@ihep.ac.cn}
\author{Sai Wang}
\email{wangsai@ihep.ac.cn}
\author{Zhe Chang}
\email{changz@ihep.ac.cn}
\affiliation{Institute of High Energy Physics,\\
Theoretical Physics Center for Science Facilities,\\
Chinese Academy of Sciences, 100049 Beijing, China}

\begin{abstract}
We present a scalar perturbation for the $\Lambda$CDM model, which breaks the isotropic symmetry of the universe. Based on the Union2 data, the least-$\chi^2$ fit of the scalar perturbed $\Lambda$CDM model shows that the universe has a preferred direction $(l,b)=(287^\circ\pm25^\circ,11^\circ\pm22^\circ)$. The magnitude of scalar perturbation is about $-2.3\times10^{-5}$. The scalar perturbation for the $\Lambda$CDM model implies a peculiar velocity, which is perpendicular to the radial direction. We show that the maximum peculiar velocities at redshift $z=0.15$ and $z=0.015$ equal to $73\pm28 \rm km\cdot s^{-1}$ and $1099\pm427 \rm km\cdot s^{-1}$, respectively. They are compatible with the constraints on peculiar velocity given by Planck Collaboration.
\end{abstract}

\maketitle
\section{Introduction}
The standard cosmological model, i.e., the $\Lambda$CDM model \cite{Sahni,Padmanabhan} has been well established. It is consistent with several precise astronomical observations that involve Wilkinson Microwave Anisotropy Probe (WMAP) \cite{Komatsu}, Planck satellite \cite{Planck1}, Supernovae Cosmology Project \cite{Suzuki}. One of the most important and basic assumptions of the $\Lambda$CDM model states that the universe is homogeneous and isotropic on large scales. However, such a principle faces challenges \cite{Perivolaropoulos}. The Union2 SnIa data hint that the universe has a preferred direction $(l,b)=(309^\circ,18^\circ)$ in galactic coordinate system \cite{Antoniou}. Toward this direction, the universe has the maximum expansion velocity. Astronomical observations \cite{Watkins} found that the dipole moment of the peculiar velocity field on the direction $(l,b)=(287^\circ\pm9^\circ,8^\circ\pm6^\circ)$ in the scale of $50h^{-1}\rm Mpc$ has a magnitude $407\pm81 \rm km\cdot s^{-1}$. This peculiar velocity is much larger than the value $110 \rm km\cdot s^{-1}$ given by WMAP5 \cite{WMAP5}. The recent released data of Planck Collaboration show deviations from isotropy with a level of significance ($\sim3\sigma$) \cite{Planck2}. Planck Collaboration confirms asymmetry of the power spectrums between two preferred opposite hemispheres. These facts hint that the universe may have a preferred direction.

Many models have been proposed to resolve the asymmetric anomaly of the astronomical observations. An incomplete and succinct list includes: an imperfect fluid dark energy \cite{Koivisto1}, local void scenario \cite{Alexander,Garcia}, noncommutative spacetime effect \cite{Akofor}, anisotropic curvature in cosmology \cite{Koivisto2}, and Finsler gravity scenario \cite{Chang}.

In this paper, we present a scalar perturbation for the flat Friedmann-Robertson-Walker (FRW) metric \cite{Weinberg}. Based on the Union2 data, the least-$\chi^2$ fit of the scalar perturbed $\Lambda$CDM model shows that the universe has a preferred direction. In the scalar perturbed $\Lambda$CDM model, the universe could be treated as a perfect fluid approximately. In comoving frame, however, the fluid has a small velocity $v$. It could be regarded as the peculiar velocity of the universe. The data of Planck Collaboration gives severe constraints on the peculiar velocity \cite{Planck3}. For the bulk flow of Local Group, it should be less than $254 \rm km \cdot s^{-1}$. For bulk flow of galaxy clusters at $z=0.15$, it should be less than $800 \rm km\cdot s^{-1}$.

The paper is organized as follows. In Sec. II, we present a scalar perturbation for the FRW metric. Explicit relation between luminosity and redshift is obtained. In Sec. III, we show a least-$\chi^2$ fit of the scalar perturbed $\Lambda$CDM model to the Union2 SnIa data. The preferred direction is found $(l,b)=(287^\circ\pm25^\circ,11^\circ\pm22^\circ)$. The magnitude of the scalar perturbation is at the scale of $10^{-5}$. This perturbation implies a peculiar velocity with value $73\pm28 \rm km\cdot s^{-1}$ at $z=0.15$, and $1099\pm427 \rm km\cdot s^{-1}$ at $z=0.015$. The conclusions and remarks are given in Sec. IV.

\section{Scalar perturbation for FRW metric}
The FRW metric describes the homogeneous and isotropic universe. In order to describe the deviation from isotropy, we try to add a scalar perturbation for the FRW metric. The scalar perturbed FRW metric is of the form
\begin{equation}\label{scalar perturbation}
ds^2=(1-2\phi(\vec{x}))dt^2-a^2(t)(1+2\phi(\vec{x}))\delta_{ij}dx^i dx^j.
\end{equation}
It should be noticed that the scalar perturbation field $\phi(\vec{x})$ is time-independent. And the scalar perturbation can be interpreted as a sort of space-dependent spatial curvature. By setting the scale factor $a(t)=1$, one can find that the spatial Ricci tensor of metric (\ref{scalar perturbation}) is of the form
\begin{equation}
R_{ij}=-\delta_{ij}\phi_{,k,k}.
\end{equation}
The nonvanishing components of Einstein tensor for the metric (\ref{scalar perturbation}) are given as
\begin{eqnarray}
G^0_0&=&3(1+2\phi)H^2-2a^{-2}\phi_{,i,i}~,\\
G^i_j&=&\delta_{ij}(1+2\phi)\left(H^2+2\frac{\ddot{a}}{a}\right)~,\\
G^0_j&=&-2H\phi_{,j}~,
\end{eqnarray}
where the commas denote the derivatives with respect to $x^i$, the dot denotes the derivatives with respect to cosmic time $t$ and $H\equiv\frac{\dot{a}}{a}$. The scalar perturbation breaks homogeneity and isotropy of the universe. Since $\phi$ is a perturbation, the cosmic inventory could be treated as a perfect fluid approximately. In comoving frame, however, the fluid has a perturbed velocity $v$. The energy-momentum tensor is given by
\begin{equation}
T^{\mu\nu}=(\rho+p)U^\mu U^\nu-p g^{\mu\nu} ,
\end{equation}
where $\rho$ and $p$ are the energy density and pressure
density of the fluid, respectively. Here, we set $U^\mu$ as $U^0=1,\frac{U^i}{U^0}\equiv v^i$, to first order in $v$. In this paper, we just investigate low redshift region of the universe, where the universe is dominated by matter and dark energy. Thus, the nonvanishing components of energy-momentum tensor are given as
\begin{eqnarray}
T^0_0&=&\rho_m+\rho_{de}~,\\
T^0_i&=&\rho_mv_i~,\\
T^i_j&=&\delta^i_j\rho_{de}~,
\end{eqnarray}
where $\rho_m$ and $\rho_{de}$ denote the energy density of matter and dark energy, respectively.
Then, the Einstein field equation $G^\mu_\nu=8\pi GT^\mu_\nu$ gives three independent equations
\begin{eqnarray}\label{eq 00}
(1+2\phi)H^2-\frac{2a^{-2}}{3}\phi_{,i,i}&=&\frac{8\pi G}{3}(\rho_m+\rho_{de})~,\\
\label{eq ij}
(1+2\phi)(H^2+2\frac{\ddot{a}}{a})&=&8\pi G\rho_{de}~,\\
\label{eq 0j}
H\phi_{,j}&=&-4\pi G\rho_m v_j~.
\end{eqnarray}

The energy-momentum conservation equation reads
\begin{equation}\label{conservation eq}
\frac{\partial T^\mu_\nu}{\partial x^\mu}+\Gamma^\mu_{\alpha\mu}T^\alpha_\nu-\Gamma^\alpha_{\nu\mu}T^\mu_\alpha=0,
\end{equation}
where $\Gamma^\mu_{\alpha\mu}$ is the Christoffel symbol. Then, following the theory of general relativity, we obtain the specific form of energy-momentum conservation equation for matter and dark energy in the perturbed FRW universe (\ref{scalar perturbation}). It is as follows:
\begin{eqnarray}\label{conser eq1}
\frac{\partial \rho_m}{\partial t}+3H\rho_m+\frac{\partial \rho_m v^i}{\partial x^i}&=&0,\\
\label{conser eq2}
\frac{\partial \rho_m v_i}{\partial t}+3H\rho_m v_i-\phi_{,i}\rho_m&=&0,\\
\label{conser eq3}
\frac{\partial \rho_{de}}{\partial t}&=&0,\\
\label{conser eq4}
\frac{\partial \rho_{de}}{\partial x^i}&=&0.\\
\end{eqnarray}
The equations (\ref{conser eq3}) and (\ref{conser eq4}) show that the energy density of dark energy remaining constant in our model. By making use of the field equation (\ref{eq 0j}), we find from equation (\ref{conser eq1}) that
\begin{equation}\label{continue eq}
\frac{\partial (\rho_m a^3)}{\partial t}=-aH\frac{\phi_{,i,i}}{4\pi G}.
\end{equation}
The solution of equation (\ref{continue eq}) reads
\begin{equation}\label{continue eq1}
\rho_m a^3=-\frac{\phi_{,i,i}}{4\pi G}(a-1)+\rho_{m0},
\end{equation}
where $\rho_{m0}$ denotes the energy density of matter at present. We have already used the initial condition that the present energy density of matter is constant to deduce continuity equation (\ref{continue eq1}).

The light propagation satisfies $ds=0$, which gives
\begin{equation}\label{light propa}
\frac{dt}{a(t)}=(1+2\phi(\vec{x}))\delta_{ij}dx^i dx^j~.
\end{equation}
The right-hand side of the equation (\ref{light propa}) is time-independent. During a very short time, the location of a galaxy is unchanged. Then, we get
\begin{equation}
\int^{t_0}_{t_1}\frac{dt}{a(t)}=\int^{t_0+\delta t_0}_{t_1+\delta t_1}\frac{dt}{a(t)}~.
\end{equation}
Thus, the redshift $z$ of galaxy satisfies
\begin{equation}\label{redshift}
1+z=\frac{1}{a},
\end{equation}
where we have set the scale factor $a(t)$ to be 1 at present.

A particular form of $\phi(\vec{x})$ is necessary for deriving relation between luminosity distance and redshift. The form of $\phi(\vec{x})$ is determined by the perturbed energy-momentum tensor. However, the information of the perturbed energy-momentum tensor is unknown. On the contrary, we choose a specific form of $\phi$ to determine the perturbed energy-momentum tensor. It is given as
\begin{equation}\label{phi1}
\phi=A\cos\theta~,
\end{equation}
where $A$ is a dimensionless parameter and $\theta$ is the angle between $\vec{r}$ and $z$-axis. By making use of (\ref{phi1}), we reduce the equation (\ref{eq 00}) to
\begin{equation}\label{eq 001}
\frac{da}{dt}=H_0a(1-A\cos\theta)\sqrt{\Omega_{m0}a^{-3}+1-\Omega_{m0}-\frac{4A\cos\theta}{3r^2H_0^2}a^{-3}},
\end{equation}
where $H_0$ is Hubble constant and $\Omega_{m0}\equiv8\pi G\rho_{m0}/(3H_0^2)$ is the energy density parameter for matter at present.

Combining the equations (\ref{light propa}), (\ref{redshift}), (\ref{phi1}), (\ref{eq 001}) and (\ref{continue eq1}), and using the definition of luminosity distance \cite{Weinberg}, we obtain the relation between luminosity distance and redshift
\begin{equation}\label{lumin red1}
H_0d_L=(1+z)\int_0^z\frac{(1-A\cos\theta)dx}{\sqrt{\Omega_{m0}(1+x)^3+1-\Omega_{m0}-\frac{4A\cos\theta(1+x)^5}{3H_0^2d^2_{L0}}}}~,
\end{equation}
where $d_{L0}\equiv(1+z)\int_0^z\frac{dx}{H_0\sqrt{\Omega_{m0}(1+x)^3+1-\Omega_{m0}}}$.


\section{Numerical results}
Our numerical studies are based on the Union2 SnIa data \cite{Amanullah}. Our goal is to find whether the universe has a preferred direction or not. We perform a least-$\chi^2$ fit to the Union2 SnIa data
\begin{equation}
\chi^2\equiv\sum^{557}_{i=1}\frac{(\mu_{th}-\mu_{obs})^2}{\sigma_\mu^2},
\end{equation}
where $\mu_{th}$ is theoretical distance modulus given by
\begin{equation}
\mu_{th}=5\log_{10}\frac{d_L}{\rm Mpc}+25~.
\end{equation}
$\mu_{obs}$ and $\sigma_\mu$, given by the Union2 SnIa data, denote the observed values of the distance modulus and the  measurement errors, respectively. The least-$\chi^2$ fit of the $\Lambda$CDM model gives $\Omega_{m0}=0.27\pm0.02$ and $H_0=70.00\pm0.35$ $\rm km\cdot s^{-1}\cdot Mpc^{-1}$. Before using our model to fit the Union2 SnIa data, we fix the values of $\Omega_m$ and $H_0$ as their mean values. Such an approach is valid for the scalar perturbed model, since it is just a perturbation for the $\Lambda$CDM model. Then, the least-$\chi^2$ fit of the formula (\ref{lumin red1}) gives $A=(-2.34\pm0.91)\times10^{-5}$ and $(l,b)=(287^\circ\pm25^\circ,11^\circ\pm22^\circ)$. The preferred direction is plotted as point G of Fig.1. The preferred directions given by other models are plotted in Fig.1 for contrast. Kogut {\it et al}. \cite{Kogut1993} got $(l,b)=(276^\circ\pm3^\circ,30^\circ\pm3^\circ)$ is shown as point A, Antoniou {\it et al}. \cite{Antoniou} got $(l,b)=({309^\circ}^{+23^\circ}_{-3^\circ},{18^\circ}^{+11^\circ}_{-10^\circ})$ is shown as point B, Cai and Tuo \cite{Cai and Tuo2012} got $(l,b)=({314^\circ}^{+20^\circ}_{-13^\circ},{28^\circ}^{+11^\circ}_{-33^\circ})$ is shown as point C, Kalus {\it et al}. \cite{Kalus:2013zu} got $(l,b)=({325^\circ},{-19^\circ})$ is shown as point D, Cai {\it et al}. \cite{Cai:2013lja} got $(l,b)=({306^\circ},{-13^\circ})$ is shown as point E, Chang {\it et al}. \cite{Chang} got $(l,b)=(304^\circ\pm43^\circ,-27^\circ\pm13^\circ)$ is shown as point F. Within a level of significance ($1\sigma$), it is shown in Fig.2 that our results are consistent with the one of Kogut {\it et al}. \cite{Kogut1993}, Antoniou {\it et al}. \cite{Antoniou} and Cai {\it et al}. \cite{Cai and Tuo2012}.

\begin{figure}
\includegraphics[scale=0.7]{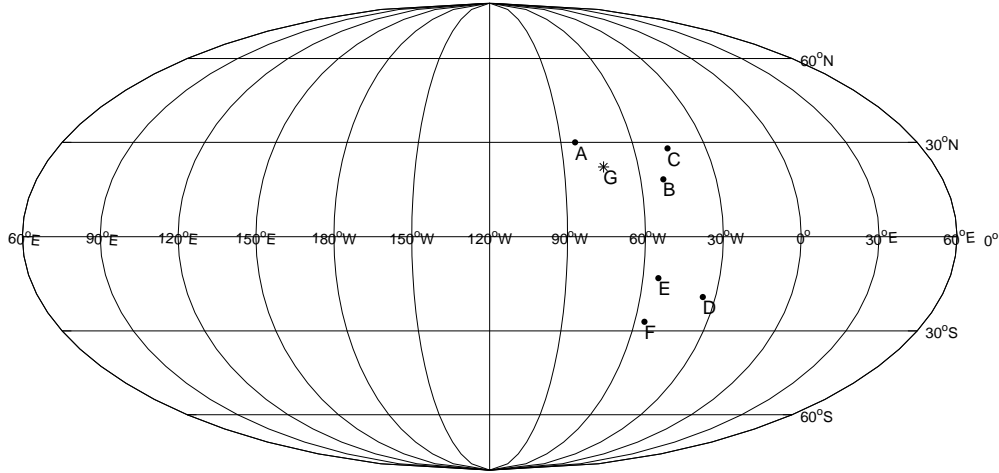}
\caption{The direction of preferred axis in galactic coordinate. The point G denotes our result, namely, $(l,b)=(287^\circ\pm25^\circ,11^\circ\pm22^\circ)$, which is obtained by fixing the parameters $\Omega_m=0.27$ and $H_0=70.00$ and doing the least-$\chi^2$ to the Union2 data for formula (\ref{lumin red1}). The results for preferred direction in other models are presented for contrast. Point A denotes the result of Kogut {\it et al}. \cite{Kogut1993}, point B denotes the result of Antoniou {\it et al}. \cite{Antoniou}, point C denotes the result of Cai and Tuo \cite{Cai and Tuo2012}, point D denotes the result of Kalus {\it et al}. \cite{Kalus:2013zu}, point E denotes the result of Cai {\it et al}. \cite{Cai:2013lja}, point F denotes the result of Chang {\it et al}. \cite{Chang}.}
\end{figure}

\begin{figure}
\includegraphics[scale=0.7]{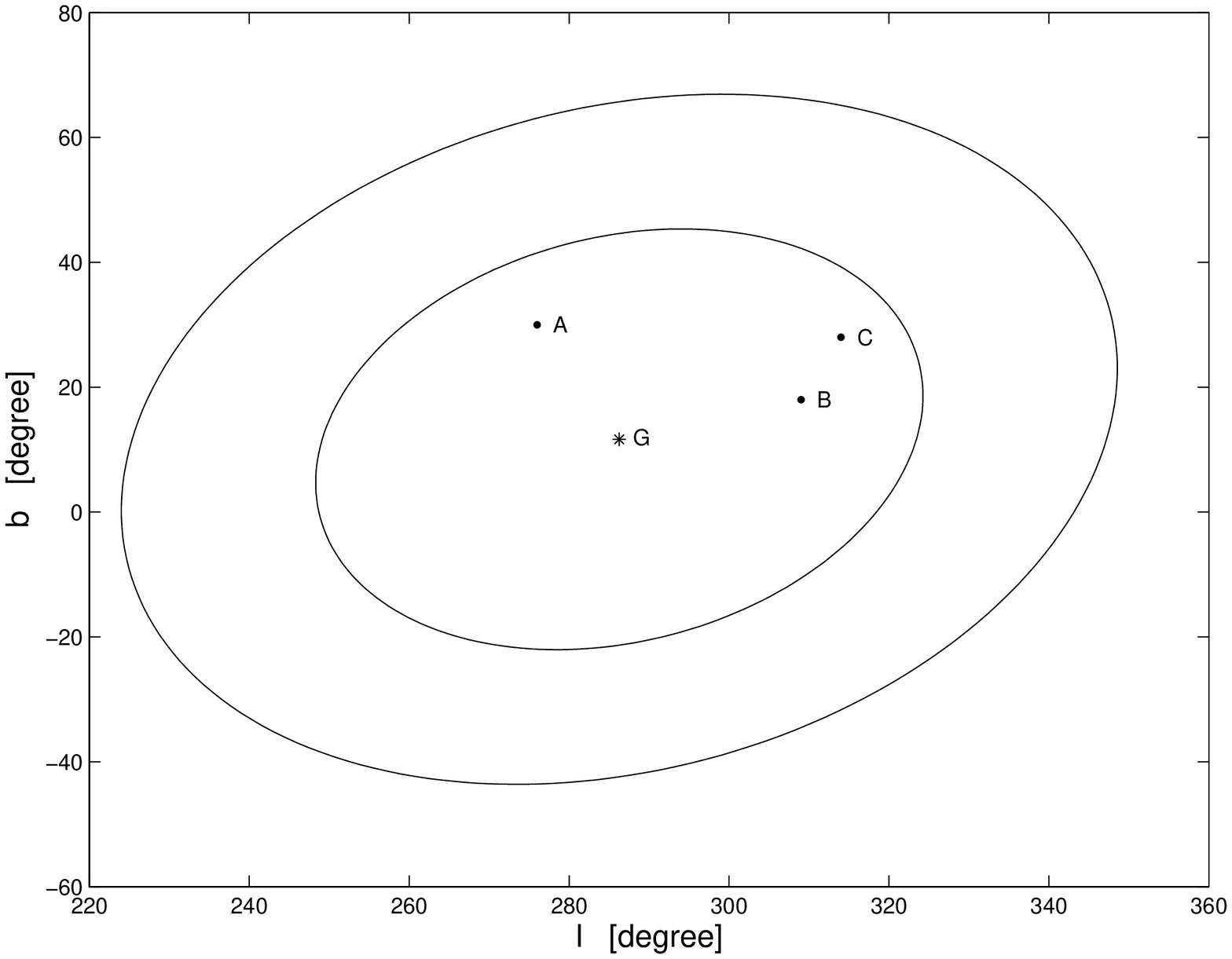}
\caption{Contours figure for the preferred direction. The contours enclose 68\% and 95\% confidence regions of the scalar perturbed $\Lambda$CDM model.}
\end{figure}

The scalar perturbation not only breaks the isotropy symmetry of the universe but also gives a peculiar velocity for the matter. By setting $\phi$ to be the form of (\ref{phi1}), we find from (\ref{eq 0j}) that
\begin{equation}\label{peculiar ve}
v\equiv\sqrt{|v_i v^i|}=a^4\frac{2H|A|\sin\theta}{3rH_0^2\Omega_{m0}}=\frac{2H|A|\sin\theta}{3H_0^2d_L\Omega_{m0}}(1+z)^{-3},
\end{equation}
where we have used the relation $d_L=(1+z)r$ to obtain the second equation. Substituting the value of $H_0d_L$ and $H$ (given by the $\Lambda$CDM model) into formula (\ref{peculiar ve}), and setting $\sin\theta=1$ (the velocity $v$ is perpendicular to the preferred direction $(l,b)=(287^\circ\pm25^\circ,11^\circ\pm22^\circ)$), we could obtain the value of peculiar velocity $v$ for a given redshift. At $z=0.15$, we get $v|_{z=0.15}\simeq73\pm28 \rm km\cdot s^{-1}$. This peculiar velocity is compatible with the result of Planck Collaboration \cite{Planck3}. it gives a upper limit $800\rm km\cdot s^{-1}$ for peculiar velocity at $z=0.15$. It should be noticed that the peculiar velocity $v$ grows with time. We should check that if the peculiar velocity at the lowest redshift in the Union2 SnIa data is compatible with the result of Planck Collaboration \cite{Planck3}. The lowest redshift in the Union2 SnIa data is 0.015. At $z=0.015$, we get $v|_{z=0.015}\simeq1099\pm427 \rm km\cdot s^{-1}$. Planck Collaboration \cite{Planck3} gives an upper limit on the bulk flow for Local Group, which equals to $254\rm km\cdot s^{-1}$. Though, our result for peculiar velocity at $z=0.015$ larger than $254\rm km\cdot s^{-1}$, with a level of significance ($1\sigma$), it is still compatible with the upper limit $800\rm km\cdot s^{-1}$ given by Planck Collaboration. And our result for peculiar velocity at $z=0.015$ represents the upper limit of peculiar velocity in the scalar perturbed $\Lambda$CDM model.

\section{Conclusions and remarks}
We presented a scalar perturbation for the $\Lambda$CDM model that breaks the isotropic symmetry of the universe. Setting the scalar perturbation of the form $\phi=A\cos\theta$, we obtained a modified relation (\ref{lumin red1}) between luminosity distance and redshift. The least-$\chi^2$ fit to the Union2 SnIa data showed that the universe has a preferred direction $(l,b)=(287^\circ\pm25^\circ,11^\circ\pm22^\circ)$, which is close to the results of Kogut {\it et al}. and Antoniou {\it et al}. and Cai {\it et al}. \cite{Kogut1993,Antoniou,Cai and Tuo2012}. Also, the least-$\chi^2$ fit to the Union2 SnIa data showed that the magnitude of scalar perturbation $A$ equals to $(-2.34\pm0.91)\times10^{-5}$. The scalar perturbation has the same magnitude with the level of CMB anisotropy. The CMB anisotropy is a possible reason for the preferred direction of the universe.

The peculiar velocity was obtained directly from the Einstein equation (\ref{eq 0j}). The numerical calculations showed that the peculiar $v|_{z=0.15}\simeq73\pm28 \rm km\cdot s^{-1}$ and $v|_{z=0.015}\simeq1099\pm427 \rm km\cdot s^{-1}$. They are compatible with the results of Planck Collaboration \cite{Planck3}. It should be noticed that the peculiar velocity we obtained is perpendicular to the radial direction.

Bianchi cosmology \cite{Rosquist} has been studied for many years. It admits a set of anisotropic metrics such as Kasner metric \cite{Misner}. The three dimensional space of Bianchi cosmology admits a set of Killing vectors $\xi_i^{(a)}$ which obey the following property
\begin{equation}
\left(\frac{\partial \xi_i^{(c)}}{\partial x^k}-\frac{\partial \xi_k^{(c)}}{\partial x^i}\right)\xi_{(a)}^i\xi_{(b)}^k=C^c_{ab},
\end{equation}
where $C^c_{ab}$ is the structure constant of the symmetry group of the space. The scalar perturbation field $\phi(\vec{x})$ completely destroys the rotational symmetry of cosmic space. It means that no Killing vectors corresponding to the symmetry group of three dimensional cosmic space. Thus, there is no obvious relation between the Bianchi cosmology and our model.
\vspace{1cm}
\begin{acknowledgments}
We would like to thank  Y. G. Jiang for useful discussions. Project 11375203 and 11305181 supported by NSFC.
\end{acknowledgments}


\begin{thebibliography}{999}
\bibitem{Sahni}V. Sahni, Class. Quant. Grav. {\bf 19}, 3435 (2002).
\bibitem{Padmanabhan}T. Padmanabhan, Phys. Rept. {\bf 380}, 235 (2003).
\bibitem{Komatsu}E. Komatsu, {\it et al}. (WMAP Collaboration), Astrophys. J. Suppl. {\bf 192}, 18 (2011).
\bibitem{Planck1}Planck Collaboration, arXiv:1303.5062.
\bibitem{Suzuki}N. Suzuki, {\it et al}., Astrophys. J. {\bf 746}, 85 (2012).
\bibitem{Perivolaropoulos}L. Perivolaropoulos, arXiv:1104.0539.
\bibitem{Antoniou}I. Antoniou and L. Perivolaropoulos, JCAP {\bf 1012}, 012 (2012).
\bibitem{Watkins}R. Watkins, H. A. Feldman and M. J. Hudson, Mon. Not. Roy. Astron. Soc. {\bf 392}, 743 (2009).
\bibitem{WMAP5}G. Hinshaw, {\it et al}., Astrophys. J. Suppl. {\bf 180}, 225 (2009).
\bibitem{Planck2}Planck Collaboration, arXiv:1303.5083.
\bibitem{Koivisto1}T. Koivisto and D. F. Mota, Phys. Rev. D {\bf 73}, 083502 (2006).
\bibitem{Alexander}S. Alexander, T. Biswas, A. Notari and D. Vaid, JCAP {\bf 0909}, 025 (2009).
\bibitem{Garcia}J. Garcia-Bellido and T. Haugboelle, JCAP {\bf 0804}, 003 (2008).
\bibitem{Akofor}E. Akofor, {\it et al}., JHEP {\bf 0805}, 092 (2008).
\bibitem{Koivisto2}T. S. Koivisto, D. F. Mota, M. Quartin and T. G. Zlosnik, Phys. Rev. D {\bf 83}, 023509 (2011).
\bibitem{Chang}Z. Chang, M.-H. Li, X. Li and S. Wang, Eur. Phys. J. C {\bf 73}, 2459 (2013).
\bibitem{Weinberg}S. Weinberg, {\it Gravitation and Cosmology: Principles and Applications of the General Theory of Relativity}, John Wiley \& Sons, New York, 1972.
\bibitem{Planck3}Planck Collaboration, arXiv:1303.5090.
\bibitem{Perlmutter}S. Perlmutter, {\it et al}., Astrophys. J. {\bf 517}, 565 (1999); A. G. Riess, {\it et al}., Astron. J. {\bf 116}, 1009 (1998); Astron. J. 117, 707 (1999).
\bibitem{Amanullah}R. Amanullah, {\it et al}., Astrophys. J. {\bf 716}, 712 (2010).
\bibitem{Kogut1993}A. Kogut, {\it et al}., Astrophys. J. {\bf 419}, 1 (1993).
\bibitem{Cai and Tuo2012}R.-G. Cai and Z.-L. Tuo, J. Cosmol. Astropart. Phys. {\bf 1202}, 004 (2012).
\bibitem{Kalus:2013zu}B. Kalus, {\it et al}., Astron. Astrophys.  {\bf 553}, A56 (2013).
\bibitem{Cai:2013lja}R. G. Cai, Y. Z. Ma, B. Tang and Z. L. Tuo, Phys. Rev. D {\bf 87}, 123522 (2013).
\bibitem{Rosquist}K. Rosquist and R. T. Jantzen, Phys. Rept. {\bf 166}, 89 (1988).
\bibitem{Misner}C. W. Misner, K. S. Thorne and J. A. Wheeler, {\it Gravitation}, W. H. Freeman and Company, San Francisco, 1973.



\end{thebibliography}
\end{document}